\documentclass[preprint,12pt,3p]{elsarticle}
\usepackage[utf8]{inputenc} 
\usepackage[T1]{fontenc}    
\usepackage{hyperref}       
\usepackage{url}            
\usepackage{booktabs}       
\usepackage{amsfonts}       
\usepackage{nicefrac}       
\usepackage{microtype}      
\usepackage{lipsum}
\usepackage{multirow}
\usepackage{amsmath,amssymb,graphicx}
\usepackage{float} 
\usepackage[dvipsnames]{xcolor}
\usepackage{relsize}
\usepackage{soul} 
\newcommand{\bigvarepsilon}{\mathlarger{\mathlarger{\varepsilon}}}
\newcommand{\bigsigma}{\mathlarger{\mathlarger{\sigma}}}

\usepackage{siunitx}
\usepackage{array}
\usepackage{lmodern}
\usepackage{color}

\usepackage[normalem]{ulem} 

\journal{Elsevier}
\begin{document}
\begin{frontmatter}
\title{Mechanical properties of Tetragraphene single-layer: A Molecular Dynamics Study}
\author[UFPI]{Wjefferson H. S. Brandão}
\author[UFPI]{Acrisio L. Aguiar\corref{authorA}}
\cortext[authorA]{Corresponding authors}
\ead{acrisiolins@ufpi.edu.br}
\author[Unicamp1]{Alexandre F. Fonseca}
\author[Unicamp1]{D. S. Galvão}
\author[IFPI,Unicamp1]{J. M. De Sousa\corref{authorA}}
\ead{josemoreiradesousa@ifpi.edu.br}

\affiliation[UFPI]{organization={Departamento de Física, Universidade Federal do Piauí},
            addressline={Ininga}, 
            city={Teresina},
            postcode={64049-550}, 
            state={Piauí},
            country={Brasil}}


\affiliation[Unicamp1]{organization={Departamento de Física Aplicada, Instituto de Física “Gleb Wataghin”, Universidade Estadual de Campinas},
            addressline={Rua Sérgio Buarque de Holanda, 777 - Cidade Universitária}, 
            city={Campinas},
            postcode={13083-859}, 
            state={São Paulo},
            country={Brasil}}
\affiliation[IFPI]{organization={Instituto Federal de Educação, Ciência e Tecnologia do  Piauí},
            addressline={Primavera}, 
            city={São Raimundo Nonato},
            postcode={64770-000}, 
            state={Piauí},
            country={Brasil}}
            

\begin{abstract}
A quasi-2D semiconductor carbon allotrope called {\it tetrahexcarbon}, also named {\it tetragraphene}, was recently proposed featuring an unusual structure combining squared and hexagonal rings. Mechanical and electronic properties of tetragraphene have been predicted based on first-principles Density Functional Theory (DFT) calculations. However, a comprehensive study of its mechanical behavior under different temperatures is still lacking. In this work, using fully atomistic reactive molecular dynamics (MD) simulations, we investigate the mechanical properties of monolayer tetragraphene under tensile strain from the linear regime up to the complete structural failure (fracture). Different temperatures were considered and the results were compared to that of two other known planar carbon alotropes: graphene and penta-graphene. 
One interesting result is that tetragraphene experiences a transition from crystalline to an amorphous structure by either temperature or tension application. At room temperature, the critical strains along the two orthogonal unit-cell directions of tetragraphene are $38$\% and $30$\%, which is higher than that for graphene and penta-graphene. Tetragraphene Young's modulus values along its directions are from three to six times smaller than that of graphene and about 57\% that of penta-graphene at room temperature. Ultimate tensile strength values along the two directions of tetragraphene were obtained and also shown to be smaller than that of graphene and penta-graphene. 
\end{abstract}

\begin{keyword}
Reactive Molecular Dynamics \sep Mechanical Properties  \sep Tetragraphene \sep Nanofracture pattern.
\end{keyword}

\end{frontmatter}
\section{Introduction}
\label{INT}

Two-dimensional (2D) nanostructures have received much attention after the isolation of a single layer of graphite (graphene) in 2004~\cite{novoselov2004electric}. 
Its single-atom thickness made it to become the new paradigm of flat or quasi-flat materials, and the combination of special properties like lightness, flexibility, conductivity, and resistance~\cite{fuchs2008introduction,wei2012extraordinary,neto2009electronic,abergel2010properties,choi2010effects,ovid2013mechanical,zhang2005experimental,novoselov2005two} opened up a wide range of potential applications from fiber production to plasmonics and super-capacitors~\cite{cheng2014graphene,berry2013impermeability,grigorenko2012graphene,stoller2008graphene}. Although having all these special properties, graphene null bandgap imposes limitations to electronic applications, which lead to the search of new 2D materials that, at least, share some of the graphene properties. Many promising structures have, then, been proposed and/or experimentally realized. In particular, graphene has many 2D carbon allotropes, such as graphynes~\cite{baughman1987structure}, penta-graphene~\cite{zhang2015penta}, phagraphene~\cite{phag2015} and popgraphene~\cite{wang2018popgraphene} among others. Here, a new member of this family of carbon planar structures is going to be studied using classical molecular dynamics methods.

Amongst the above examples, penta-graphene is not a 'true' one-atom-thick structure because of its structural buckling. It is, then, commonly considered a ``quasi-2D'' one. The combination of $sp^2$ and $sp^3$ bonds in penta-graphene generates a planar pentagonal network of carbon atoms of about 1.2 \AA\ thickness. It is thermally stable semiconductor with an indirect bandgap value of $3.25$ eV. Also, its Young's modulus and Poisson's ratio values were predicted to be $263.8$ GPa.nm and $-0.068$, respectively~\cite{zhang2015penta}. Further Density Functional Theory (DFT) and molecular dynamics (MD) studies confirmed the negative nature of Poisson's ratio and the values of Young's modulus of penta-graphene membranes ranging from 200 to 330 GPa.nm
\cite{de2021computational,brandao2021penta,de2018mechanical}. One interesting result is that when tensile strained, the structure of penta-graphene is predicted to suffer a transition from the pentagons to planar structures having 6, 7 or even 8 carbon rings depending on the conditions~\cite{de2021computational,cranford2016-pentatransition,rahaman2017}. 

Recently, a related new planar carbon nanostructure called tetrahexcarbon was proposed~\cite{tetrahex}. It is quasi-2D in the same sense mentioned above for penta-graphene, and is composed of a combination of squared and hexagonal rings (Figs. \ref{fig:graf-tetrag-pentag}(d)-(f)). This combination has motivated the suggestion of another name for the structure that will be considered in this work: {\it tetragraphene}~\cite{de2019electronic}. Similar to penta-graphene, tetragraphene has also $sp^2$ and $sp^3$ carbon hybridizations and is also a semiconductor material with a direct bandgap of 3.70 eV. The high electronic mobility ($1.463\times10^4$ cm$^2$V$^{-1}$s$^{-1}$) is notable in this material that has smaller cohesion energy than penta-graphene and might be a good option for applications in high-performance electronic devices~\cite{tetrahex}. Another, more recent study of the electronic properties of tetragraphene, showed that it presents metallic and semiconductor characteristics, with bandgap values in the range of 0.82 eV - 4.4 eV, depending on structural parameters~\cite{de2019electronic}. The mechanical properties of tetragraphene have been also investigated based on DFT calculations~\cite{tetrahex,de2019electronic,QunPRApplied2020,TetraHidrogenetaed2020}. Young's modulus along \textit{zigzag} (x) and \textit{armchair} (y) directions of tetagraphene were predicted to be in the range of 280 and 288 GPa.nm, and shear modulus of 165 GPa.nm. Wei {\it et al.}~\cite{QunPRApplied2020}, in particular, showed that tetragraphene presents intrinsic negative Poisson's ratio and predicted rupture stresses and strains, along zigzag and armchair directions of the structure, between 40 and 50 GPa.nm, and between 33 and 40 \%, respectively. Some studies also revealed interesting elastic properties of tetragraphene under hydrogen and fluorine functionalizations ~\cite{TetraHidrogenetaed2020,kilic2020}. 
However, to our knowledge, no comprehensive and detailed study of the tetragraphene mechanical properties has been reported at the level of classical MD simulations, and it is one of the objectives of the present study.

In this work, the mechanical and structural properties of a single tetragraphene monolayer, beyond the elastic regime, including its fracture dynamics, at different temperature values are investigated. As benchmarks for comparisons, the same properties of penta-graphene and graphene structures, obtained from the same MD potential and protocols of simulations, are presented. We have carried out fully atomistic reactive (ReaxFF \cite{mueller2010development,van2001reaxff}) molecular dynamics (MD) simulations using the well-known LAMMPS code \cite{plimpton1995fast}. It is important to investigate the mechanical and structural properties of tetragraphene at different physical conditions in order to investigate possible structural transitions. Our results show that monolayers of tetragraphene exhibit critical strains between three and four times higher than graphene depending on the temperature and direction. On the other hand, the Young's modulus and Ultimate Tensile Strength (UTS) of tetragraphene are smaller than that of graphene. Different from penta-graphene, which was shown to give rise to other 2D crystal structures when tension-strained~\cite{de2021computational,cranford2016-pentatransition,rahaman2017}, we have observed a tension and temperature-induced transition in tetragraphene from crystalline to amorphous structure before total fracture. At 1000 K, the tetragraphene structure spontaneously evolves to an amorphous one without the application of stress. After the recent synthesis and characterization of an amorphous 2D carbon structure~\cite{Toh2020,Felix2020}, studies of routes to produce amorphous carbon can be of interest.  Young's modulus and UTS values of tetragraphene are, on average, smaller than that of penta-graphene, whereas its critical strain is greater than that of penta-graphene~\cite{brandao2021penta}. 

In section $2$, we present the computational methodology and detailed information about the structure of tetragraphene. In section $3$, we present the results for the mechanical properties of tetragraphene. The results for the Young's Modulus, Ultimate Tensile Strength (UTS) and critical strain ($\bigvarepsilon_{C}$) of tetra-graphene were compared to that of graphene and penta-graphene. In section $4$, the conclusions and main remarks of this work are presented. 

\section{Computational Method}
\label{CM}
\subsection{Reactive Molecular Dynamics Simulation Method}

\begin{figure*}[!htb]
\centering
\includegraphics[scale=0.6]{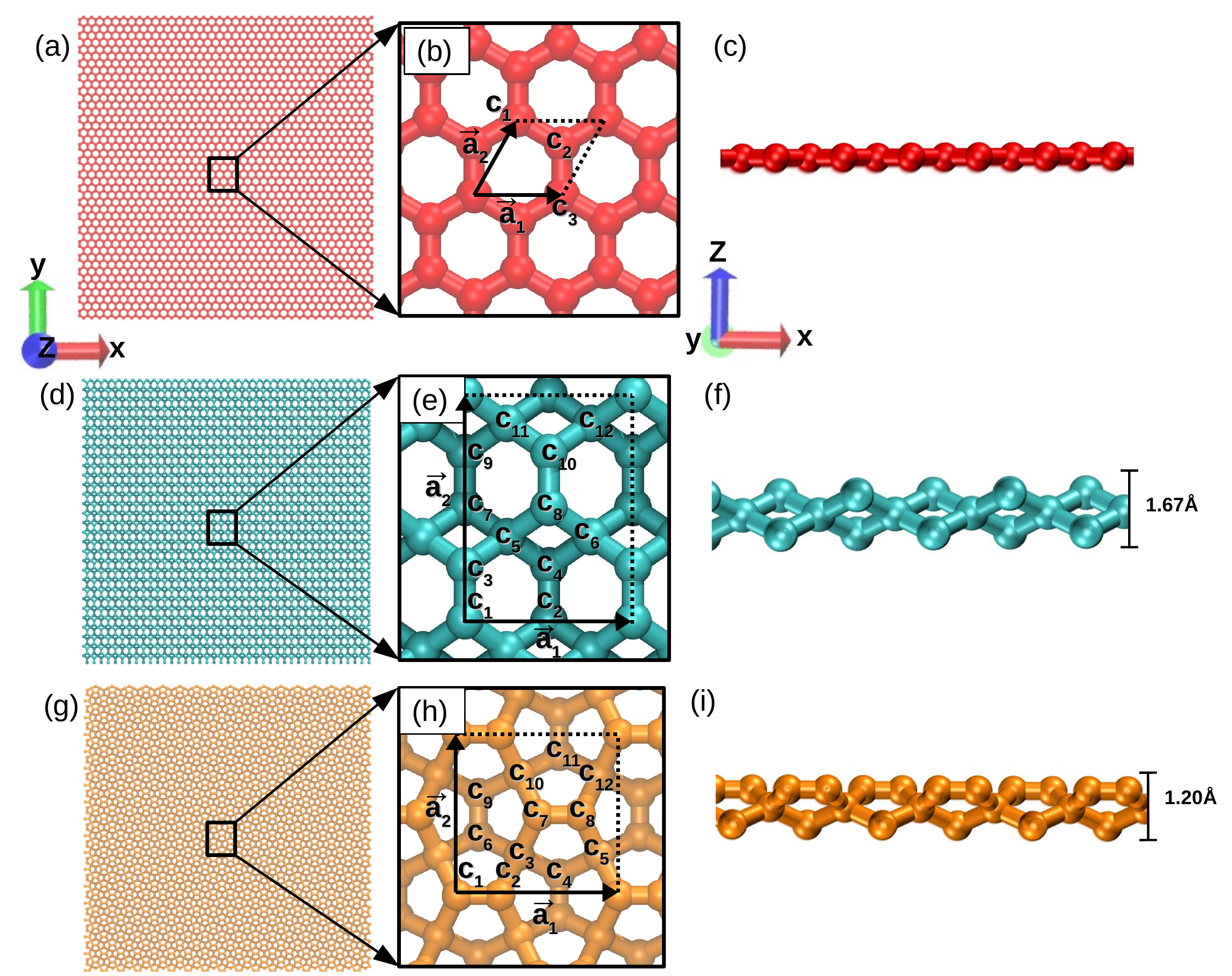}
\caption{ Atomistic structures of graphene (first row), tetragraphene (middle row) and penta-graphene (bottom row). Each colum shows supercell samples of the structures (first column, (a) (d) (g)); zoomed regions of the structures with the corresponding unit cell drawn within dotted line with its atoms named for reference in the text and unit vectors (second column (b) (e) (h)); and the lateral views showing the thickness of the structures (third column (c) (f) (i)). For tetragraphene the bond distance values are C$_1$-C$_3$=1.40 \AA\ and C$_3$-C$_5$=1.535 \AA, the lattice vectors $\boldsymbol{a}_1$=4.53 \AA\ $\hat{x}$ and $\boldsymbol{a}_2$=6.11 Å $\hat{y}$. The $x$ and $y$ directions correspond to zigzag and armchiar directions in all structures, respectively.}
\label{fig:graf-tetrag-pentag}
\end{figure*} 

The atomic configurations of graphene, tetragraphene, and penta-graphene monolayers, including their corresponding unit cells, thickness, and carbon-carbon bond-length values, are shown in Fig. \ref{fig:graf-tetrag-pentag}. While graphene structure is formed by an arrangement of pure sp$^2$ 
carbon atoms (see Fig. \ref{fig:graf-tetrag-pentag}(a)-(c)), tetragraphene is composed of a mixture of $sp^2$ and $sp^3$ ones. The two types of C-C carbon bonds (parallel to $y$-axis in Fig. \ref{fig:graf-tetrag-pentag}(d)-(f)) are not in the same plane, which results that tetragraphene is a non-single-atom thickness membrane. The structure of tetragraphene has similar features to that of penta-graphene, which is also shown for comparison in panels (g)-(i) of Fig. \ref{fig:graf-tetrag-pentag}. The optimized values of bond lengths and bond angles are shown in Table \ref{tab:bonds-angles} for comparison. Theses structural parameters of tetragraphene are in good agreement with those obtained with DFT in the literature~\cite{de2019electronic,hosseinian2017dft}.

The stress-strain protocols of simulations (described in detail below) were also performed with graphene, penta-graphene, and tetragraphene in order to verify and validate the methods and have a benchmark to which the results of tetragraphene can be compared~\cite{de2021computational,brandao2021penta}.

\begin{table}[htb!]
\centering
\caption{Bond length and angle values for graphene (G), tetragraphene (TG), and penta-graphene (PG) unit cells. See Fig. \ref{fig:graf-tetrag-pentag} and refs.\cite{tetrahex,de2019electronic,zhang2015penta} for more information.}
\begin{tabular}{|c|c|c||c|c|}
\hline
 Membrane & Bonds & Length (\AA) & Angles$\measuredangle$ & Degree ($^\circ$) \\ \hline
G & C$_1$-C$_2$ & 1.42 & C$_1$-C$_2$-C$_3$ & 120 \\ \hline \hline
\multicolumn{1}{|c|}{\multirow{5}{*}{TG}} & \multirow{2}{*}{C$_1$-C$_3$} & \multirow{2}{*}{1.48} & C$_1$-C$_3$C$_5$ & 123 \\ \cline{4-5} 
\multicolumn{1}{|c|}{} &  &  & C$_5$-C$_4$-C$_6$ & 95 \\ \cline{2-5} 
\multicolumn{1}{|c|}{} & \multirow{3}{*}{C$_3$-C$_5$} & \multirow{3}{*}{1.50} & C$_3$-C$_5$-C$_7$ & 85 \\ \cline{4-5} 
\multicolumn{1}{|c|}{} &  &  & C$_3$-C$_5$-C$_4$ & 112 \\ \cline{4-5} 
\multicolumn{1}{|c|}{} &  &  & C$_3$-C$_5$-C$_8$ & 134 \\ \hline \hline
\multicolumn{1}{|c|}{\multirow{4}{*}{PG}} & \multirow{2}{*}{C$_1$-C$_2$} & \multirow{2}{*}{1.34} & C$_1$-C$_2$-C$_3$ & 114 \\ \cline{4-5} 
\multicolumn{1}{|c|}{} &  &  & C$_3$-C$_7$-C$_{10}$ & 112 \\ \cline{2-5} 
\multicolumn{1}{|c|}{} & \multirow{2}{*}{C$_2$-C$_3$} & \multirow{2}{*}{1.55} & C$_2$-C$_3$-C$_6$ & 99 \\ \cline{4-5} 
\multicolumn{1}{|c|}{} &  &  & C$_2$-C$_3$-C$_7$ & 134 \\ \hline
\end{tabular}
\label{tab:bonds-angles}
\end{table}

All the simulations were carried out through fully atomistic MD simulations \cite{rapaport2004art,allen2017computer}. The equations of motion of all carbon atoms were integrated using the LAMMPS package~\cite{plimpton1995fast}. The reactive force field (ReaxFF) interatomic potential for carbon~\cite{mueller2010development,van2001reaxff} was used here. ReaxFF allows for dynamical and continuous chemical bond formation and dissociation, which was validated by experimental results on the heat of formation of carbonaceous systems~\cite{van2001reaxff}. The ReaxFF has been used in many successfull theoretical investigations of the mechanical and other properties of carbon and other different nanostructured systems~\cite{de2016mechanical,de2016torsional,de2019elastic,de2021computational,brandaomechanical,nair2011minimal,kocsis2014confinement}. 

A supercell of tetragraphene monolayer of $90 \times 90$ \AA\ size, with $3600$ carbon atoms, was considered here.  Young’s Modulus, UTS, and critical strain values of tetragraphene were estimated from MD simulations of tensile strain. To evaluate the temperature effects on these mechanical properties of tetragraphene, the applied strain simulations were performed at the following temperatures: $10$ K, $300$ K, $600$ K, and $1000$ K. 

Initially, structural minimizations at $0$ K of our systems were performed in order to remove any residual stress from their initial atomic configurations. After that, the membranes were tension-free thermalized at $10$ K, $300$ K, $600$ K and $1000$ K, at NPT ensemble conditions (fixed number of atoms, $N$, pressure, $P$, and temperature, $T$). Nosé-Hoover thermostat and barostat at a null pressure were used to equilibrate the structures at target $T$ values.
Tensile stress/strain calculations were, then, performed fixing $N$ and $T$, but changing the length of the system until the complete fracture process. 
Temporal evolution of carbon atoms position was obtained by the integration of Nosé-Hoover style non-Hamiltonian equations of motion, which control the initial oscillations of non-equilibrium classical reactive MD simulations~\cite{evans1985nose}. $Tdamp=100$ ($Pdamp=1000$) timesteps were used to keep the average system temperature (pressure) fixed at the desired temperature (pressure) values \cite{sutmann2002classical}. A constant engineering tensile strain rate of $\eta = 10^{-6}$ fs$^{-1}$ was used, where $L(t)=L_{0} + L_{0}\eta t$ and $L_{0}$ is the initial linear length of monolayer (which is about $90$ \AA\ for both systems). The engineering tensile strain rate is small enough for stabilization and allows carbon-carbon bond reconstructions during the stretching dynamics.
The virial stress along the stretched monolayer along $x$ and $y$-directions, is defined as \cite{subramaniyan2008continuum,garcia2010bioinspired,
buehler2008atomistic}:

\begin{eqnarray}
\sigma_{xx} = \frac{1}{\Xi}\left[ \sum_{l=1}^{N} \left( m_{l}v_{lx} + r_{lx}F_{lx} \right) \right] ,
\end{eqnarray}
and
\begin{eqnarray}
\sigma_{yy} = \frac{1}{\Xi}\left[ \sum_{l=1}^{N} \left( m_{l}v_{ly} + r_{ly}F_{ly} \right) \right] ,
\end{eqnarray}
where $\Xi$ is the monolayer area, $N$ the number of carbon atoms, $m_{l}$, $v_{lx}$, $v_{ly}$, $r_{lx}$, and $r_{ly}$ are the mass and the $x$ and $y$ coordinates of velocity and position vectors of the $l$th carbon atom of the system, respectively. The $F_{lx}$ and $F_{ly}$ are the forces per carbon atom along the $x$ and $y$ directions, respectively. From a linear fitting of the stress/strain curves at low strain, usually the linear elastic regime is satisfied for strain $< 1$\%, the Young's Modulus ($Y_{modulus}$) can be estimated by~\cite{buehler2008atomistic,landau1986theory}:

\begin{eqnarray}
Y_{modulus}^{i} =  \frac{\sigma_{ii}}{\varepsilon_{i}},
\label{eq:ymodx}
\end{eqnarray}
where index $i$ refers to $x$ or $y$ direction,  $\varepsilon_{i} = (L_i - L_{i0})/L_{i0}$ 
is the load strain applied to the monolayers along the $i=x$ or $i=y$ directions, and $\sigma_{ii}$ ($ii$ = $xx$ or $yy$) is the tensor component of the virial stress along the $x$ or $y$ directions, respectively, of the monolayers.

In order to analyze the C-C bond stretching behavior during the stress/strain dynamics of the membranes, we calculated the spatial distribution of von Mises stress, which is defined as \cite{buehler2008atomistic,landau1986theory}:
\begin{eqnarray}
\sigma_{VonMises}^{i} = \sqrt{\frac{(\sigma_{xx}^{i}-\sigma_{yy}^{i})^{2} + 
(\sigma_{yy}^{i}-\sigma_{zz}^{i})^{2} + (\sigma_{xx}^{i}-\sigma_{zz}^{i})^{2} +
6 \left[(\sigma_{xy}^{i})^{2} + (\sigma_{yz}^{i})^{2}  + (\sigma_{zy}^{i})^{2}\right]}{2}} .
\label{Eq4}
\end{eqnarray} 
Therefore, the average of spatial atomic stress distribution for all structures under the stretching process can be monitored by using the von Mises stress tensor. $\sigma_{xx}$, $\sigma_{yy}$, $\sigma_{zz}$ are the uniaxial tensor components, which are parallel ($x$ and $y$) and perpendicular ($z$) to the applied load strain. $\sigma_{xy}$, $\sigma_{yz}$ and $\sigma_{zx}$ are shear stress components \cite{buehler2008atomistic,landau1986theory}. This stress/strain methodology has been successfully used in previous studies in the literature, such as the twist of carbon nanotubes \cite{de2016torsional} and the mechanical properties of  one-~\cite{de2018mechanical,de2019elastic,de2021mechanical,brandaomechanical} or two-dimensional~\cite{de2021computational,de2018mechanical,de2016mechanical,pereira2020temperature} carbon structures. This method have been also used in  theoretical investigations of nanostructures under high strain rate\cite{de2016carbonSCROLLS,woellner2018structural,bizao2018scale,de2020carbonPEAPODS}, including its presence in environmental gaseous atmospheres \cite{autreto2013dynamics,autreto2014site,de2020hydrogenation}.

In this work, the following properties were obtained from the stress-strain curves: Young's modulus, defined by Eqs. \eqref{eq:ymodx}; the UTS (Ultimate Tensile Strength), which is the highest stress value reached (different from the critical breaking stress value $\bigsigma_C$); and the critical strain of rupture, $\bigvarepsilon_C$.

\section{Results and Discussions}
\subsection{Thermal stability}

Initially, in order to validate our methodology, we compared our findings with those of graphene and penta-graphene data from our previous works\cite{brandao2021penta,de2021nanostructures}. 

\begin{figure}[htb!]
    \centering
    \includegraphics[scale=0.55]{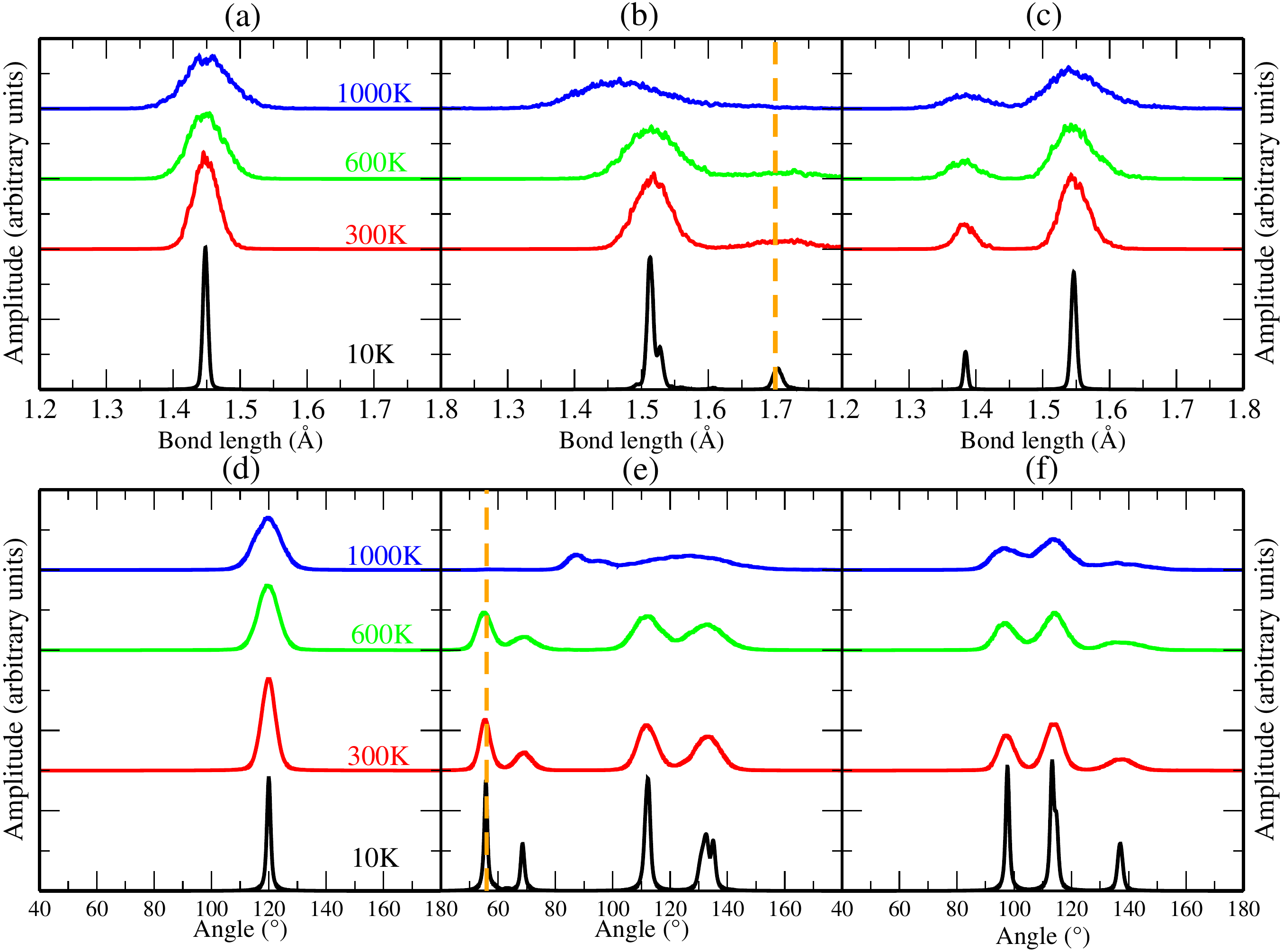}
    \caption{Bond length and angle values distribution for: graphene in (a) and (d); tetragraphene in (b) and (e); and penta-graphene in (c) and (f), respectively. Data information for graphene and penta-graphene provided with the permission of J. M. de Sousa\cite{de2021nanostructures} and W. H. S. Brandão \textit{et. al}\cite{brandao2021penta}.
    }
    \label{fig:ang-bond-dist}
\end{figure}

We present in Fig. \ref{fig:ang-bond-dist} the distribution of bond length (Fig. \ref{fig:ang-bond-dist}(a)-(c)) and bond angle (Fig. \ref{fig:ang-bond-dist}(d)-(f)) for graphene, tetragraphene and penta-graphene at the final stage of thermalization at $10$ K, $300$ K, $600$ K, and $1000$ K. The characteristic bond lengths for graphene at 10 K were obtained to be averaged in 1.45 \AA\ (black curve in Fig. \ref{fig:ang-bond-dist}(a)) with a broadening of the peaks increasing with temperature. 
A slight displacement of the C-C peaks to higher values at 1000 K (see Fig. \ref{fig:ang-bond-dist}(a)), is consistent with thermal elongation of the C-C bonds\cite{mosterio2014}. Tetragraphene has three characteristic C-C bond length peaks at 10 K: 1.51\AA, 1.52\AA\ and 1.70\AA\ (Fig. \ref{fig:ang-bond-dist}(b)) which, except for the last value, are consistent with fluctuations of the C$_1$--C$_3$ and C$_3$--C$_5$ bond lengths (see Fig. \ref{fig:graf-tetrag-pentag}(e) and Table \ref{tab:bonds-angles}). The small peak at 1.70\AA\ at Fig. \ref{fig:ang-bond-dist}(b) (orange vertical dashed line) corresponds to off-plane movements of the carbon atoms of the square diagonal (distance between non-bonded C$_4$ and C$_8$ atoms, see Fig. \ref{fig:graf-tetrag-pentag}(e)). This C--C distance is too large to be considered as a covalent bond and will be not considered as such.
From 300 K up to 1000 K, the peak around 1.70\AA\ spreads out, and as expected there is an increase of the width of C-C peak centered around 1.51\AA, due to a thermal rearrangement of the membrane. In fact, at 1000 K, we observe a strong structural rearrangement of tetragraphene membrane which is characterized by losing its crystallinity and nominal thickness (cf. Fig. \ref{fig:graf-tetrag-pentag}(e)). The lack of crystallinity can be observed by the large broadening of the peak and its shift 
to a lower value (\hspace{0.2mm}1.46 \AA), which is
characteristic for pure $sp^2$ covalent carbon materials. One should note that the thermal stability of penta-graphene has been recently studied, using a similar methodology~\cite{brandao2021penta}. Penta-graphene also presents two C--C bond length peaks at 1.38 \AA\ and 1.55 \AA, that broaden under the increase of temperature.

Analysis of the bond angle distribution is another way to verify changes within the structures. The bond angle distribution for graphene reveals its excellent thermal stability, with peaks centered at 120$^\circ$ up to a temperature of 1000 K (Fig. \ref{fig:ang-bond-dist}(d)). Penta-graphene presents three peaks for the angle distribution, 97.70$^\circ$ , 113.36$^\circ$ and 137.01$^\circ$. The last one is weak at 600 K and disappears for temperaturesabove 600 K. The other two peaks start to get mixed at 1000 K, indicating its structure thermal instability (see Fig. \ref{fig:ang-bond-dist}). Tetragraphene presents good thermal stability but up to 600 K.  For this case (Fig. \ref{fig:ang-bond-dist}(e)), the angle distribution at 10 K clearly distinguishes five peaks located at $\sim$ 56$^\circ$, $\sim$ 69$^\circ$, $\sim$ 113$^\circ$, $\sim$ 132$^\circ$ and $\sim$ 134$^\circ$. The first angle, $\sim$ 56$^\circ$ (orange vertical dashed line in Fig. \ref{fig:ang-bond-dist}(e)) is formed by the covalent C$_5$-C$_8$ bond and a line formed by non-bonded for C$_8$ and C$_4$ atoms (see Fig. \ref{fig:graf-tetrag-pentag}(e)), so it should not be considered. The other four values of angles correspond to C$_3$--C$_5$--C$_7$, C$_1$--C$_3$--C$_5$, C$_3$--C$_5$--C$_8$ and C$_3$--C$_5$--C$_4$ pairs of bonds, respectively. Also as expected, due to thermal vibrations of the structure, all the peaks spread out and considerably reduce their intensity. At 300 and 600 K, we could only identify four main angle peaks since those at $\sim$ 132$^\circ$ and $\sim$ 134$^\circ$ overlapped. In fact, these peaks correspond to C$_3$--C$_5$--C$_8$ and C$_3$--C$_5$--C$_4$ pairs of bonds that, from Fig. \ref{fig:graf-tetrag-pentag}(e), are expected to present similar values. At 1000 K, the characteristic peak distribution of tetragraphene is not seen anymore. In fact, at 1000 K the angle distribution gets spread out along a wide range of angles, which indicates 
that a kind of thermal amorphization happened. These results show that tetragraphene structure is not stable at high temperature, which is also corroborated by its C--C bond length distribution at 1000 K (see Fig. \ref{fig:ang-bond-dist}(b)). Recently, based on \textit{ab initio} molecular dynamics (AIMD) simulations, Babu {\it et al}.~\cite{tetrahex} argued that
tetragraphene is thermally stable at 300 K, and that no significant C--C bond reconstructions happen to the structure up to 1000 K (except for a large off-plane membrane bending). However, 
as our calculations use five times larger tetragraphene supercells, which provides more room for
the atoms of the tetragraphene membrane to reconfigure themselves to smaller energy configurations, 
our results are expected to be more realistic.  

\subsection{Elastic properties and fracture mechanism}

We now focused on the results from the reactive classical MD simulations on the elastic response of the structures under uniaxial strain. The obtained differences between graphene, penta-graphene, and tetragraphene become evident when we analyze both representative MD snapshot frames and the stress-strain curves. 
Here, in the main manuscript, we will present and discuss the atomic configurations of the tetragraphene structure during the load strain applied along its zigzag and armchair directions at 300 K. 
Results for other temperature values (600 K and 1000 K) will be summarized but can be found, in detail, in Supplementary Material (Figs. S1 up to S4). 

\begin{figure*}[htb!]
\centering
\includegraphics[width=\linewidth]{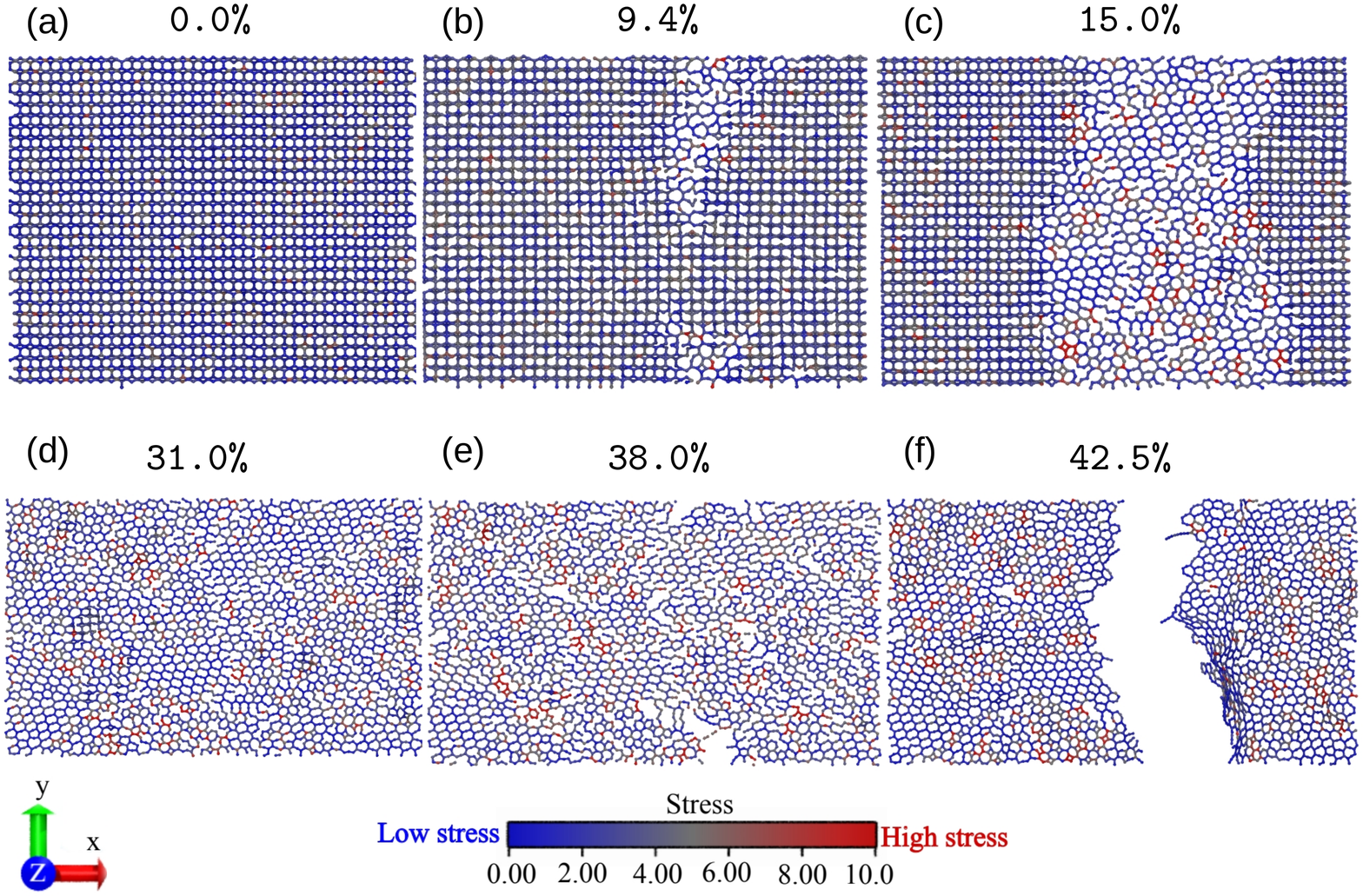}
\caption{Representative MD snapshots of the tetragraphene structure subjected to different strain levels along its zigzag direction, at room temperature (300K): (a) 0.0\% non-stretched sheet, (b) $9.4$\%, a structure just after the beginning of the amorphization process, (c) $15.0$\%, a structure showing an intermediate stage of the amorphization process, (d) $31.0$\%, a structure where the amorphization process is close to the end, (e) $38.0$\%, a structure where the C-C bonds start breaking in the amorphous phase, and (f) $42.5$\%, fully fractured structure. The color bar at the bottom indicates the stress concentration within the monolayer where blue (red) stands for low-stress (high-stress) levels, respectively.}
\label{fig:tg-strain-x}
\end{figure*} 

\begin{figure*}[htb!]
\centering
\includegraphics[width=\linewidth]{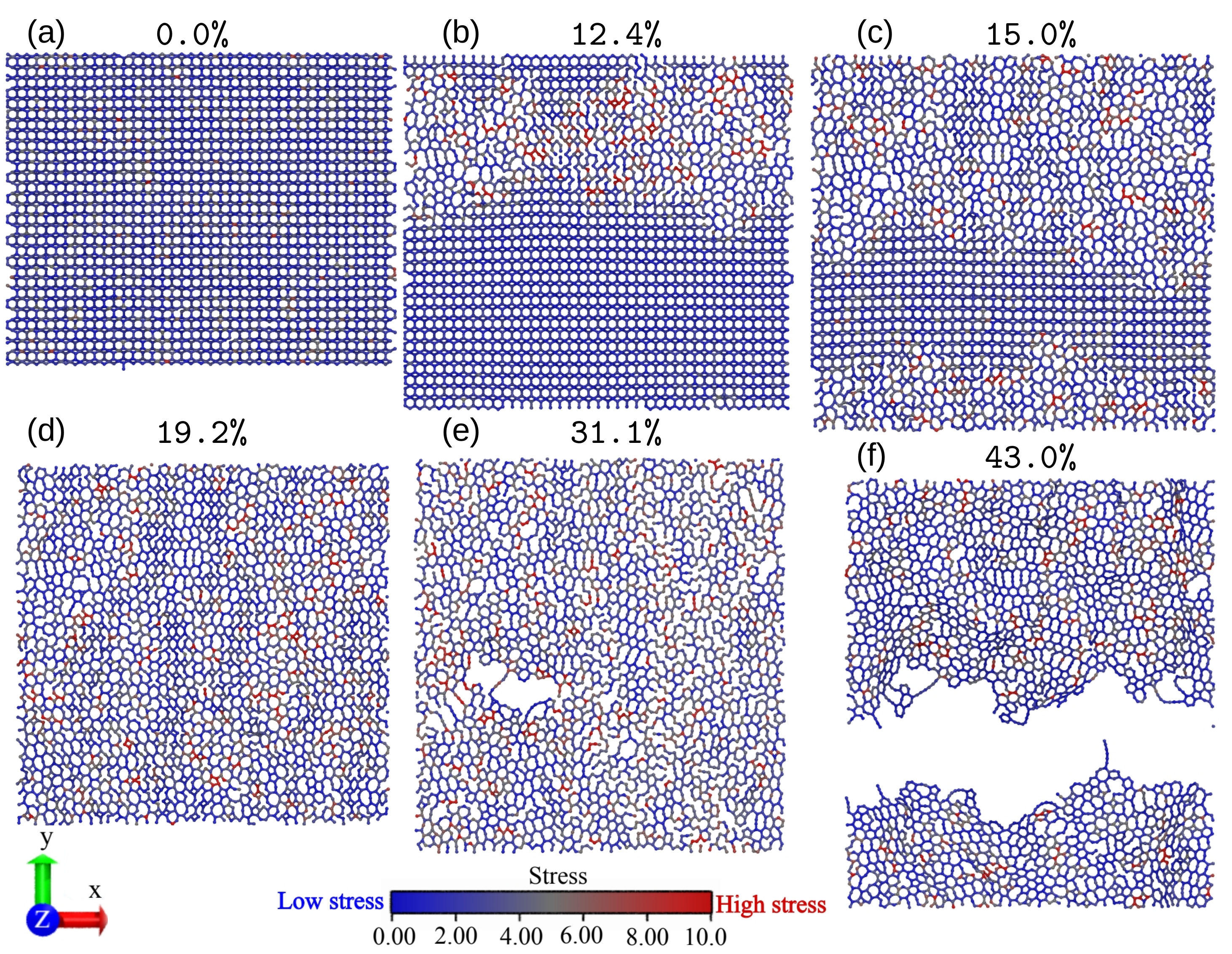}
\caption{Representative MD snapshots of the tetragraphene structure subjected to different strain levels along its armchair-direction, at room temperature (300K):
(a) $0$\%, non-stretched sheet, in (b) $12.4$\%, a structure just after the beginning of the amorphization process, (c) $15.0$\%, a structure showing an intermediate phase of the amorphization process, (d) $19.2$\%, a structure at the very end of the amorphization process, (e) 31.1\% a structure after the C-C bonds start breaking in the amorphous phase and (f) 43.0\%, a fully fractured structure. The color bar at the bottom indicates the stress level in the monolayer where blue (red) stands for low-stress (high-stress) levels, respectively.}
\label{fig:tg-strain-y}
\end{figure*} 

Representative MD snapshots for different load strain values can be seen in Figs. \ref{fig:tg-strain-x} and \ref{fig:tg-strain-y} for tetragraphene zigzag and armchair-directions, respectively. They show a completely different fracture process when compared to that of graphene\cite{de2021nanostructures}. While in graphene, C--C bond breaking precedes its fracture, tetragraphene experiences, first, a structural transition, from crystalline to amorphous, after which carbon chains evolved to the fracture.
Therefore, our reactive MD simulations show that tetragraphene monolayer presents significant reconstructions well before the complete fracture, which implies a longer deformation process when compared to graphene at 300 K. 
The calculated elastic quantities (Young's modulus, critical stress, and fracture strain) have shown to be quite different from those of graphene and will be further discussed. 
In Fig. \ref{fig:tg-strain-x}, we can see representative MD snapshots of the tetragraphene monolayer under different load strain values applied along the zigzag direction at room temperature. In Fig. \ref{fig:tg-strain-x}(a), we observed a stable tetragraphene monolayer at $0$\% of strain. At $9.4$\% of strain (see Fig. \ref{fig:tg-strain-x}(b)), we observe the beginning of a transition effect over the structure, where the atoms in the lattice tend to reorganize and the material starts to become similar to graphene, with a one-atom thickness and featuring mainly distinct four-, six- and eight-membered carbon rings well before the fracture. This flattening process is only finished at about $31.0$\% of strain (cf. Fig. \ref{fig:tg-strain-x}(c)-(d)). After that, the structures holds some additional tensile strain and at $38.0$\% of strain, it starts fracturing  with stretched C--C bonds aligned along the zigzag direction (Fig. \ref{fig:tg-strain-x}(e)). The complete membrane fracture is found at $42.5$\% of strain, as can be seen in Fig. \ref{fig:tg-strain-x}(f).

Similar results were found for the armchair direction at room temperature, as can be seen in Fig.~\ref{fig:tg-strain-y}. However, in that case, the flattening transition of tetragraphene starts at about $12.4$\% and finishes at $19.2$\% of strain (see Fig.~\ref{fig:tg-strain-y}(b)-(d)). The membrane failure also starts with stretched C--C bonds aligned along the armchair direction and takes place at $31.1$\% (see Fig. \ref{fig:tg-strain-y}(e)). The complete fracture is found only at $43.0$\% of strain, with two pieces visibly separated, as observed in Fig. \ref{fig:tg-strain-y}(f). For more details of the whole stretching process of the tetragraphene monolayer at different temperatures, see supplementary material (Figs. from S1 to S4 and movies V1 and V2).

\begin{figure}[htb!]
    \centering
    \includegraphics[scale=0.65]{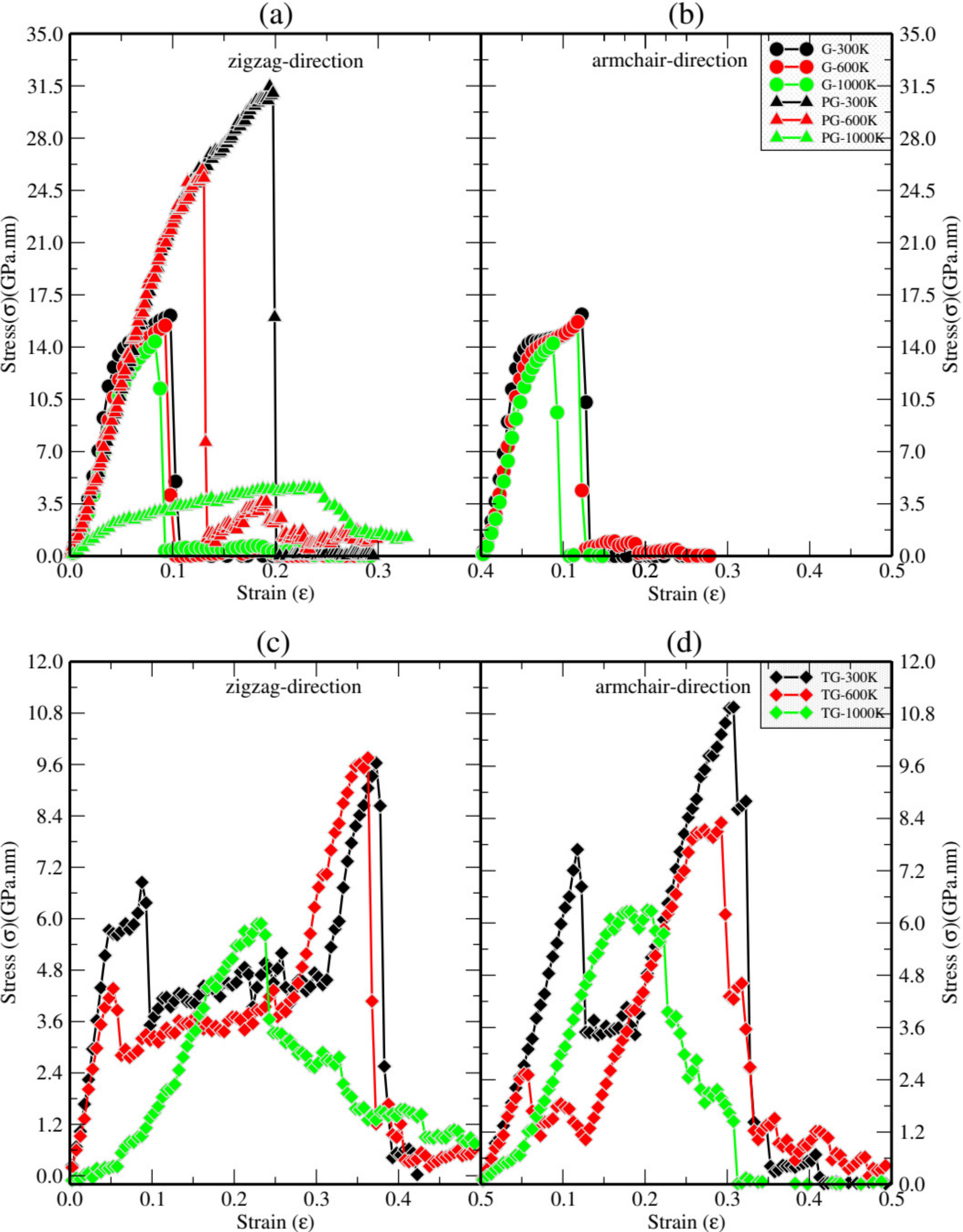}
    \caption{Stress versus strain curves for graphene (G) (a)-(b), penta-graphene (PG) (triangle up symbol in (a)), and tetragraphene (TG) (c)-(d) monolayers obtained from MD simulations with the ReaxFF potential at 300 K, 600 K, and 1000 K for the zigzag (left panels) and armchair directions (right panels), respectively. Data information for graphene and penta-graphene obtained with permission from J. M. de Sousa\cite{de2021nanostructures} and W. H. S. Brandão \textit{et. al.}\cite{brandao2021penta}, respectively.}
    \label{fig:ss-graf-tetrag}
\end{figure}

At room temperature, (see, for example, Fig. 6 of ref. \cite{de2021nanostructures}) it is known that tensioned graphene along its armchair direction 
starts failing at 12.8 \% and completely cracks at 13.2 \%. Similar results were also found along its zigzag direction. 
At 600 K (1000 K), clean fractures in graphene are observed at critical strains of 9.64\% (8.89\%) and 12.13\% (9.32\%) along the zigzag and armchair directions, respectively\cite{de2021nanostructures}. A similar effect of buckled-to-flat convergence for 2D membranes was also seen in the case of penta-graphene stretched for temperatures higher than 600K. \cite{brandao2021penta}. 

In Fig.\ref{fig:ss-graf-tetrag}, we present the obtained stress/strain curves for graphene, pentagraphene, and tetragraphene structures simulated at 300 K, 600 K, and 1000 K temperatures. Our results clearly show significant differences in stress-strain responses of those carbon membranes.
For graphene at 300 K (black circles/curves in main plot of Figs. \ref{fig:ss-graf-tetrag}a and \ref{fig:ss-graf-tetrag}b), we clearly note two regimes: first a linear regime up to $\varepsilon_{x}=\varepsilon_{y}\sim 0.05$ 
followed by a plastic-like regime up to the rupture. In the plastic-regime, graphene monolayer presents C--C bond breaking up to the fracture point, which is characterized by an abrupt stress value drop to zero at $\varepsilon_{C}\sim 0.10$ ($\varepsilon_{C}\sim 0.13$) for zigzag-direction (armchair-direction). At higher temperatures (600 K and 1000 K, red and green circles/curves, respectively), the stress/strain curves for graphene shows a small decrease in UTS e $\varepsilon_{C}$ values. This behavior for graphene was already observed for other nanostructures under thermal effects\cite{shen2010temperature,goyal2019study,lee2008measurement,cadelano2009nonlinear, shao2012temperature,pereira2020temperature}.

Penta-graphene has similar stress-strain curves compared to graphene (see the curves formed by triangle up symbols in Fig. \ref{fig:ss-graf-tetrag}a), with a critical strain value equal to 0.20 (0.13) for a temperature of 300 K (600 K), as shown in Fig. \ref{fig:ss-graf-tetrag}a. The UTS  and critical strain values decrease with increasing temperature. It is interesting to note that the temperature dependence is not observed for $Y_{modulus}$ values, but the penta-graphene structure, contrary to the graphene, is no longer stable at 1000 K, where
an overall structural degradation is observed\cite{brandao2021penta}. This structural degradation was also observed in the case of tetragraphene as we will discuss next.

In the tetragraphene case (Fig \ref{fig:ss-graf-tetrag}c and Fig. \ref{fig:ss-graf-tetrag}d), we observed several sudden stress drops during the reactive non-equilibrium MD tension simulations of the structure, which suggests an atypical response to the applied mechanical load strain for both zigzag and armchair-directions. 
At 300 K and 600 K, after the linear regime, we noted from the stress/strain curve either the begging of plastic regime or sudden stress drop (but not to zero stress). The values of strain at which these drops happen coincides with the previously mentioned amorphization process, where several carbon bond reconstructions take place and where the tetragraphene loses its original off-plane atomic configuration (cf. Fig.\ref{fig:tg-strain-x} and Fig.\ref{fig:tg-strain-y}). 
During these bonds reconstructions, the stress/strain curves are roughly flat or present a very small slope, before the final stage of additional stretching and breaking the C--C bonds. At this final stage, the stress/strain curves suddenly increase (which means hardening of the amorphous carbon membrane) until finally dropping to zero. In other words, our MD results suggest that tetragraphene monolayer has a peculiar transition in its nanostructural atomic configuration, from the original configuration (quasi-2D material featuring a combination of square and hexagonal rings) to a flattened 2D amorphous carbon form, followed by the cracking for strain values higher than $30$\%. 

It is interesting to notice that the stress-strain curves for tetragraphene at 300 K show a significant stress increasing after $\varepsilon_{x}\sim 0.31$ and $\varepsilon_{y}\sim 0.19$, which is related to the end of the flattening transition of the tetragraphene membrane. Thus, the amorphous structure is stable over a strain range higher for the zigzag-direction than for the armchair one (ranging from 9\% to 31\% for the zigzag direction and from 12\% to 19\% for the armchair one, respectively). After that, we finally observed a significant stress drop to zero at $\varepsilon_{x}\sim 0.38$ and $\varepsilon_{y}\sim 0.30$, which is consistent with results previously discussed (cf. Fig.\ref{fig:tg-strain-x} and Fig.\ref{fig:tg-strain-y}). In general, comparing the critical strain values obtained from the stretching process of graphene and tetragraphene monolayers, we can see that the tetragraphene at 300 K stands a critical strain of almost four times larger than graphene for the zigzag direction. For the armchair direction, the tetragraphene monolayer at room temperature stands a critical strain of almost three times larger than graphene.

At 600 K, similar to what was observed for 300 K, the fast increase of the stress values just before the total fracture, now starts at  $\varepsilon_{x}\sim 0.26$ and $\varepsilon_{y}\sim 0.13$, along zigzag and armchair directions of the structure, respectively, which suggest a thermal softening of tetragraphene.

The stress-strain curves obtained for tetragraphene at 1000 K (green curves in Fig.\ref{fig:ss-graf-tetrag}c and \ref{fig:ss-graf-tetrag}d) exhibits a  completely different shape as compared to that of previous temperature values. 
We have observed that at 1000 K, the tetragraphene completely loses its characteristic structural ordering (see Figs. \ref{fig:ang-bond-dist}b and \ref{fig:ang-bond-dist}d), which is consistent with the nanomaterial amorphization. 
At 1000 K, the structure presents a slope of the stress/strain curve at low strains much smaller than that observed at lower temperatures.
Due to the structural changes of tetragraphene at 1000 K, we did not collect the values of the elastic constants at this temperature.

\begin{table}[htb!]
    \caption{Some elastic quantities for Graphene (G), Tetragraphene (TG) and Penta-graphene (PG) estimated on the linear limit of 5\%. Graphene and Penta-graphene results with permissions from  W. H. S. Brandao \textit{et. al.} and J. M. de Sousa, respectively (see refs.\cite{brandao2021penta,de2021nanostructures}). }
    \centering
    \begin{tabular}{|c|c|c|c|c|}
            \hline
         \textbf{System (direction)} & \textbf{T (K)} & $\mathbf{Y_{modulus}}$ \textbf{(GPa.nm)} & \textbf{UTS (GPa.nm)} & $\mathbf{\bigvarepsilon_C}$ \\ \hline
            \multirow{3}{*}{G (zigzag)} 
            & 300 & 314.34 $\pm$ 3.60 & 16.19 $\pm$ 0.03 & 0.10 \\ \cline{2-5}
            & 600 & 273.20 $\pm$ 3.17 & 15.63 $\pm$ 0.02 & 0.10 \\ \cline{2-5}
            & 1000& 243.87 $\pm$ 2.27 & 14.76 $\pm$ 0.04 & 0.09 \\ \hline
            \multirow{3}{*}{G (armchair)} 
            & 300 & 313.33 $\pm$ 3.66 & 16.53 $\pm$ 0.02 & 0.13 \\ \cline{2-5}
            & 600 & 273.79 $\pm$ 3.37 & 15.86 $\pm$ 0.03 & 0.12 \\ \cline{2-5}
            & 1000& 237.80 $\pm$ 2.88 & 14.52 $\pm$ 0.03 & 0.09 \\
            \hline
            \hline
         \multirow{3}{*}{TG (zigzag)} 
         & 300 & 126.08 $\pm$ 1.79 & 9.85 $\pm$ 0.06 & 0.38 \\ \cline{2-5}
         & 600 & 93.84 $\pm$ 1.81 & 10.10 $\pm$ 0.07 & 0.37 \\ \cline{2-5}
         & 1000&   - & - & - \\ \hline
         \multirow{3}{*}{TG (armchair)} 
         & 300 &  51.75 $\pm$ 0.94  & 11.19 $\pm$ 0.04 & 0.30 \\ \cline{2-5}
         & 600 &  48.72 $\pm$ 1.45  & 8.80 $\pm$ 0.14 & 0.30 \\ \cline{2-5}
         & 1000&  - & - & - \\ \hline
         \multirow{3}{*}{PG (zigzag)}
         & 300 & 218.97 $\pm$ 2.56 & 32.58 $\pm$ 0.22 & 0.20 \\ \cline{2-5}
         & 600 & 227.15 $\pm$ 3.17  & 26.94 $\pm$ 0.37 & 0.13 \\ \cline{2-5}
         & 1000 & - & - & - \\ \hline
    \end{tabular}
    \label{tab:graf-elastic}
\end{table}

We present in Table \ref{tab:graf-elastic} the results of some of the elastic quantities ($Y_{modulus}$, UTS and $\bigvarepsilon_c$) obtained from the MD simulations for graphene and tetragraphene membranes for all studied temperatures (300, 600, and 1000 K) with stress applied along both zigzag and armchair directions. Penta-graphene data\cite{brandao2021penta,de2021nanostructures} was also presented for comparison. Due to the symmetric behavior of mechanical response for Penta-graphene, only one stretching direction was chosen. Firstly, we observed that for graphene, the values of $Y_{modulus}$ and UTS do not depend significantly on the stretching direction, but there are slight differences in the critical strain, with higher values for the armchair direction, i. e., graphene has a greater elongation when stretched along the armchair direction. Furthermore, the temperature effects for graphene are reflected in the elastic constants, where temperature increasing reduces, in general, the $Y_{modulus}$ values. On the other hand, the $\bigvarepsilon_C$ values remain almost unchanged, while the UTS values have a slight decrease. It is well-known that graphene monolayers have experimentally high Young's modulus values, at about 340 GPa.nm, and considerable UTS, approximately 44 GPa.nm \cite{lee2008measurement}. It was reported from DFT calculations coupled with quasi-harmonic approximations, that the $Y_{modulus}$ for temperatures of 0, 300, and 1000 K is given by 358.77, 358.41, and 352.20 GPa.nm, respectively, and UTS of 41.63, 41.34 and 39.92 GPa.nm, for these respective temperatures~\cite{shao2012temperature}. Using the same methodology (ReaxFF potential) other authors reported $Y_{modulus}$ values for graphene at 300 K (600 K) of 332.34 (330.74) GPa.nm and 330.42 (321.15) GPa.nm for the zigzag and armchair directions, respectively~\cite{pereira2020temperature}.
Therefore, our results for $Y_{modulus}$ (Table \ref{tab:graf-elastic}) are in agreement with these values found in the literature.

Table \ref{tab:graf-elastic} shows a significant anisotropy in the $Y_{modulus}$ of the tetragraphene membrane. It seems that it is more rigid along the zigzag direction. The Young's modulus of tetragraphene for the zigzag direction (armchair-direction) is $Y_{modulus}=126.08$ GPa.nm ($Y_{modulus}=51.75$ GPa.nm) at room temperature.
Recently, Wei {\it et al.} by using  DFT calculations, have not found any significant anisotropy of $Y_{modulus}$ values\cite{QunPRApplied2020}. However, they found that tetragraphene is more ductile along the zigzag direction axis compared to the armchair one. We did not observe any significant anisotropy in the UTS values but a slight difference in $\bigvarepsilon_C$ values concerning
zigzag and armchair directions. 

$Y_{modulus}$ and UTS values found for tetragraphene are smaller than that of graphene. On the other hand, the critical strain ($\bigvarepsilon_C$) values are larger in the case of tetragraphene. Therefore, tetragraphene can be easier stretched than graphene, and requires less stress to reach the fracture regime. Comparing our findings to other $sp^3$ ``quasi-2D'' material, penta-graphene has a Young's Modulus of approximately 219 GPa.nm and 227 GPa.nm, for temperatures of 300 K and 600 K, respectively\cite{brandao2021penta} as compared to 257.6 GPa.nm obtained with DFT\cite{de2021computational}. In addition, penta-graphene has UTS values around 32 GPa.nm and 27 GPa.nm, and a critical strain of 20\% and 13\% for the same temperatures, respectively. These values represent greater structural rigidity in the case of penta-graphene. Tetragraphene, otherwise, is revealed to have greater flexibility. Similar to penta-graphene, tetragraphene also loses its natural structural crystallinity for temperatures up to 1000 K. As expected, the morphology differences between tetragraphene, pentagraphene and graphene (see Fig. \ref{fig:graf-tetrag-pentag}) will directly affect their mechanical properties.   



\section{Conclusions and Remarks}

In summary, through fully atomistic classical reactive MD simulations, we investigated the structural stability, elastic properties, and fracture patterns of a novel carbon allotrope called tetragraphene. The results were contrasted to that of graphene and penta-graphene in order to verify its advantages and disadvantages. Our  MD results show that tetragraphene at room temperature stands a critical strain of $38$\% and $30$\%, so greater than the values obtained for graphene along zigzag and armchair directions, respectively. The effect of temperature on elastic values is noticeable, reducing those values as the temperature increases. Bond length and angle value distributions allowed us to show that different from graphene, tetragraphene experiences a transition from crystalline to amorphous structure by either thermal or combined thermal/tension effects. Penta-graphene also exhibits a similar structural transition but not to an amorphous structure.  
This aspect of structural transitions under the application of external strain might have interesting applications as mechanical sensors and/or fuses and also occurs to related 1D structures as shown in a recent study about the mechanics of penta-graphene nanotubes~\cite{de2021mechanical}. Similar to graphene and penta-graphene, carbon linear chains are formed before the final states of structural fracture of tetragraphene at room and high temperatures. The $Y_{modulus}$ and UTS values obtained from our MD simulations for tetragraphene monolayer are smaller than that of graphene: the $Y_{modulus}$ is around three(six) times smaller and the UTS around 39\% (32\%) lower, for zigzag (armchair) direction at 300 K. The nanostructured topology/morphology is a determining factor for the differences on these results. Thus, we hope that this work will contribute to the increasing library of theoretical results of new 2D carbon allotropes.

\section{Acknowledgements}

This work was supported in part by the Brazilian Agencies CAPES, CNPq and FAPESP. J.M.S and D.S.G thank the Center for Computational Engineering and Sciences at Unicamp for financial support through the FAPESP/CEPID Grant \#2013/08293-7. J.M.S acknowledges CENAPAD-SP (Centro Nacional de Alto Desenpenho em São Paulo - Universidade Estadual de Campinas - UNICAMP) for computational support process (proj842). A.L.A. acknowledges CNPq (Process No. $427175/20160$) for financial support. W.H.S.B., A.L.A. and J.M.S  thank the Laboratório de Simulação Computacional Cajuína (LSCC) at Universidade Federal do Piauí for computational support. A.F.F thanks the Brazilian Agency CNPq for Grant No. 303284/2021-8 and São Paulo Research Foundation (FAPESP) for Grant No. \#2020/02044-9. This research also used the computing resources and assistance of the John
David Rogers Computing Center (CCJDR) in the Institute of Physics “Gleb Wataghin”, University of Campinas.

\newpage
\bibliographystyle{elsarticle-num}
\bibliography{TetraGraphene.bib}
\end{document}